\documentclass[preprint,3p,times]{elsarticle}

\usepackage{amssymb,amsmath,amsfonts,mathrsfs,bm,gensymb}
\usepackage{graphicx,float,subfig}
\usepackage{lineno}

\usepackage{geometry}
\usepackage{hyperref}

\hypersetup{
    colorlinks=true,
    linkcolor=black,
    citecolor=black,
    filecolor=black,
    urlcolor=black,
}

\makeatother

\biboptions{sort&compress}

\begin{document}

\begin{frontmatter}

\title{Surface wave non-reciprocity via time-modulated metamaterials}

\author[add1]{A. Palermo\corref{corr1}} 
\ead{antonio.palermo6@unibo.it}
\author[add2]{P. Celli\corref{corr1}}
\ead{paolo.celli@stonybrook.edu}
\author[add3]{B. Yousefzadeh}
\author[add4]{C. Daraio}
\author[add1]{A. Marzani}
\cortext[corr1]{Corresponding authors and equal contributors}

\address[add1]{Department of Civil, Chemical, Environmental and Materials Engineering, University of Bologna, 40136 Bologna, Italy}
\address[add2]{Department of Civil Engineering, Stony Brook University, Stony Brook, NY 11794, USA}
\address[add3]{Department of Mechanical, Industrial, and Aerospace Engineering, Concordia University, Montreal, Quebec H3G 1M8, Canada}
\address[add4]{Division of Engineering and Applied Science, California Institute of Technology, Pasadena, CA 91125, USA}

\begin{abstract}

We investigate how Rayleigh waves interact with modulated resonators located on the free surface of a semi-infinite elastic medium. We begin by studying the dynamics of a single resonator with time-modulated stiffness, we evaluate the accuracy of an analytical approximation of the resonator response and identify the parameter ranges in which its behavior remains stable. Then, we develop an analytical model to describe the interaction between surface waves and an array of resonators with spatio-temporally modulated stiffness. By combining our analytical models with full-scale numerical simulations, we demonstrate that spatio-temporal stiffness modulation of this elastic \emph{metasurface} leads to the emergence of non-reciprocal features in the Rayleigh wave spectrum.  Specifically, we show how the frequency content of a propagating signal can be filtered and converted when traveling through the modulated medium, and illustrate how surface-to-bulk wave conversion plays a role in these phenomena. Throughout this article, we indicate bounds of modulation parameters for which our theory is reliable, thus providing guidelines for future experimental studies on the topic.

\vspace{15px}
\noindent{\textbf{This article may be downloaded for personal use only. Any other use requires prior permission of the authors and Elsevier Publishing. This article appeared in}: \emph{Journal of the Mechanics and Physics of Solids} 145, 104181 (2020) \textbf{and may be found at}: \url{https://doi.org/10.1016/j.jmps.2020.104181}}
\vspace{10px}

\end{abstract}


\end{frontmatter}



\section{Introduction}
The prospect of achieving non-reciprocity in elastic systems is becoming increasingly appealing to the physics and engineering communities~\cite{nassar2020}. This is motivated by the potential exploitation of this effect to realize mechanical diodes and other uni-directional devices~\cite{Boechler2011, Maznev2013, Sklan2015, Devaux2015, Zhou2019, Brandenbourger2019}. In non-reciprocal systems, wave-like excitations propagate with markedly-different amplitudes in one direction and the opposite. One way to achieve this effect is by modulating the properties of the system in space and time~\cite{Lurie97}. The dynamic behavior of mechanical systems with time-varying parameters has attracted the scientific community for more than a century~\cite{Rayleigh87,Raman}. However, the simultaneous variation of the elastic or inertial properties of a medium in both time and space has not received much attention in the mechanics community partly due to the infeasibility of the experiments. 
Only recent advances in smart structures~\cite{Airoldi2011, Hatanaka2014, Bilal2017}, together with fundamental studies on spatio-temporally modulated periodic media~\cite{Lurie97,Swinteck2015, Trainiti2016,Nassar2017jmps, Nassar2017prsa}, have allowed the realization of such systems in the context of periodic materials.

The phenomenon of time modulation-based non-reciprocity can be effectively explained with a one-dimensional example. Consider a 1D phononic crystal generated by periodically arranging an array of unit cells. Assume that the properties of each cell (stiffness and/or mass) can be independently varied in time. If we coordinate this variation in neighboring units to generate a wave-like pattern of properties that varies in space and time, we create a pump or modulating wave. Under specific frequency and wavelength constraints, mechanical waves that propagate in this system can interact with the modulating wave. In turn, this can lead to the appearance of asymmetric Bragg scattering bandgaps located at different frequency ranges for waves propagating from left to right and from right to left, and to non-reciprocal propagation~\cite{Swinteck2015, Trainiti2016, Deymier2017, Yi2018}. In physical terms, this spatio-temporal modulation breaks time-reversal symmetry. Similar considerations apply to locally-resonant metamaterials featuring an elastic wave-carrying medium equipped with a set of auxiliary resonators~\cite{Liu2000}. In this case, a wave-like modulation of the properties of the resonators causes the appearance of additional asymmetric features within the dispersion relation, such as bandgaps and veering points~\cite{Nassar2017prsa, Nassar2017eml, Attarzadeh2018, Chen2019, Huang2019}. Exciting a modulated metamaterial at specific frequencies leads to phenomena such as non-reciprocal wave filtering and frequency conversion of transmitted/reflected waves~\cite{Nassar2017eml}.

So far, investigations on elastic wave non-reciprocity via time-modulated resonators have been limited to axial and flexural waves in either discrete phononic systems~\cite{Wang2018} or beam-like metamaterials~\cite{Chen2019, Attarzadeh2020, Marconi2020}. However, it is of interest to extend this concept to elastic waves propagating near the surface of a semi-infinite medium, also known as surface acoustic waves (SAW). In this context, metamaterials can be realized by arrays of resonators located on the free surface, and are therefore known as \emph{elastic metasurfaces}~\cite{Colquitt2017}. To the best of our knowledge, surface wave non-reciprocity has been so far demonstrated only in semi-infinite structured media with a gyroscopic architecture~\cite{Zhao2020}. Achieving surface wave non-reciprocity on elastic half-spaces via metasurfaces could lead to the realization of novel SAW devices for high-frequency applications where phononic systems have already shown their promise, from acoustifluidics and particle manipulation~\cite{Guo2015, Collins2016} to mechanical signal processing~\cite{Hatanaka2014, Cha2018Nano}. 

\begin{figure}[!htb]
\centering
\includegraphics[scale=1.0]{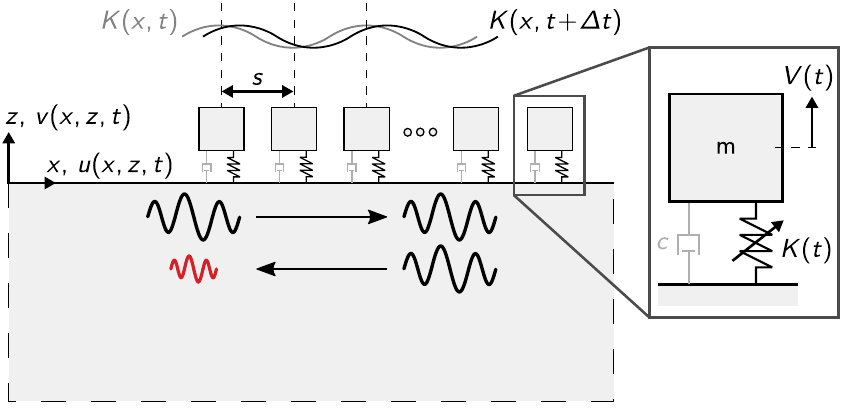}
\caption{Schematic of a time-modulated metasurface, depicting the non-reciprocal propagation of surface waves. A sinusoidal space-time evolution of the stiffness function of the resonators, $K(x,t)$, is illustrated. The inset is a close-up on one of the $N$ identical resonators placed on the free surface of the semi-infinite elastic medium.}
\label{f:met}
\end{figure}

In this work, we study how surface waves of the Rayleigh type interact with spatio-temporally modulated metasurfaces, as illustrated in the schematic in Fig.~\ref{f:met}. 
We use a combination of analytical tools and numerical simulations to investigate the effects of temporal stiffness modulations on an isolated resonator, and to identify ranges of modulation parameters where a small-modulation approximation is valid. We leverage this understanding to derive analytical solutions for the dispersion relation of Rayleigh surface waves interacting with a spatio-temporally modulated metasurface. In particular, we describe the interaction between the incident and scattered fields generated by the modulated resonators and predict the appearance of directional wave responses.
Additionally, by means of a first-order asymptotic analysis, we estimate how the modulation parameters affect the extent of the non-reciprocal wave features.
We confirm our analytical findings via numerical simulations, and demonstrate non-reciprocal wave effects such as one-way filtering and frequency conversion for transmitted and reflected signals. While our work is entirely theoretical, we envision that our analysis could guide the experimental realization of modulated metasurfaces, featuring, for example, electromechanical~\cite{Alan2019, Marconi2020} or tunable contact resonators~\cite{Palermo2019}.

The rest of the article is organized as follows. In Section~\ref{s:sdof}, we analyze the free response, stability and response to base excitation of a single time-modulated resonator. In Section~\ref{s:saw}, we study the interaction of Rayleigh waves with arrays of modulated surface resonators and obtain the dispersion curves. In Section~\ref{s:nr}, we use numerical analyses to further study the effects of spatio-temporal modulations on non-reciprocal propagation of surface waves. The conclusions and outlook of our work are reported in Section~\ref{s:concl}.

\section{Dynamics of a modulated resonator}
\label{s:sdof}

We begin by focusing on the dynamics of a single resonator. Two scenarios are captured by this analysis: a fixed, rigid substrate (Section~\ref{s:sfree}) and an oscillating, rigid substrate (Section~\ref{s:base}). These analyses allow us to better understand the interaction between the surface waves and an array of modulated resonators. By comparing analytical predictions and numerical simulations on a single resonator, we gain an understanding on the effects of stiffness modulations, we evaluate the quality of our analytical predictions, and we explore the stability of these modulated systems. This information allows us to set bounds on the choice of modulation parameters to be used for the surface wave-metasurface analysis. 

\subsection{Free vibrations}
\label{s:sfree}
\begin{figure*}[!htb]
\centering
\includegraphics[scale=1.0]{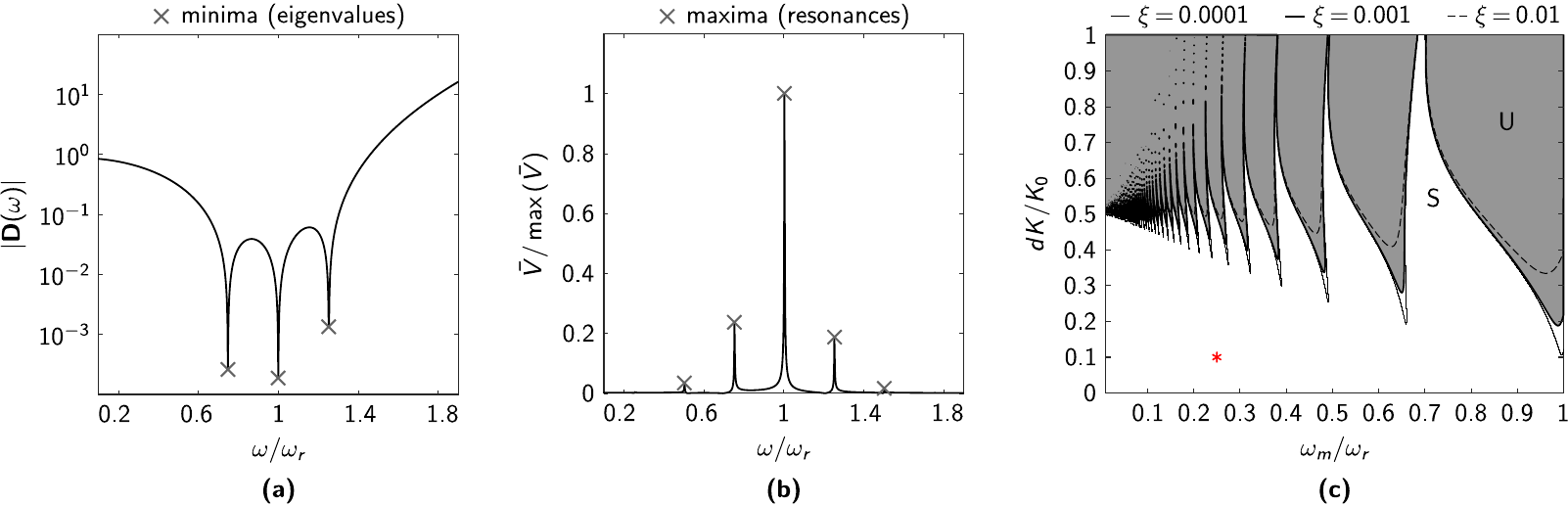}
\caption{Dynamics of a resonator with time-modulated stiffness, Eq.~\ref{e:eom}. (a) Analytical evaluation of the determinant of the dynamic matrix for $dK/K_0=0.1$, $\omega_m/\omega_r=0.25$ and $\xi=0.001$. The markers indicate the minima. (b) Fourier Transform of the response to an initial velocity for the same parameters used in (a). The markers indicate the resonance peak and its side-bands. (c) Stability diagram, as a function of the modulation parameters. The stability contours are given for three values of damping ratio $\xi$. The unstable (U) regions for $\xi=0.001$ are shaded in gray. The star marker indicates that parameters $dK/K_0=0.1$ and $\omega_m/\omega_r=0.25$ yield stable (S) results.}
\label{f:free}
\end{figure*}

We first consider a single, clamped resonator with mass $m$, damping coefficient $c$ and time-varying stiffness $K(t)$ (see the inset in Fig.~\ref{f:met}). We assume $K(t)$ to be:
\begin{equation}
K(t)=K_0+2dK \cos{\left( \omega_m t \right)},
\label{e:kdef}
\end{equation}
where $K_0$ is the average stiffness, $2dK$ is the modulation amplitude and $\omega_m$ is the modulation frequency. Note that the modulation can have the form of any periodic function~\cite{Trainiti2016,Nassar2017eml}; we choose a sinusoidal one for simplicity. For future reference, we define $\omega_r=\sqrt{K_0/m}$ and choose a small damping ratio $\xi=c/(2 m\omega_r)=0.001$. Ignoring the motion of the substrate, the equation governing the displacement $V(t)$ reads:
\begin{equation}
m\frac{d^2V}{dt^2}+c\frac{dV}{dt}+K(t)V=0.
\label{e:eom}
\end{equation}
%
This is equivalent to assuming that the substrate is fixed and rigid. {As commonly done in the literature~\cite{Vila2017, Nassar2017eml}, we assume that the restoring force exerted by the time modulated spring is obtained by multiplying stiffness and displacement at the same time instant}. Since the stiffness in Eq.~\ref{e:eom} is time-periodic, we re-write it in complex Fourier series form: 
\begin{equation}
K(t)=\sum_{p=-\infty}^{\infty}\hat{K}_p\,e^{i p \omega_m t},
\label{e:k}
\end{equation}
with Fourier coefficients defined as:
\begin{equation}
\hat{K}_p=\frac{\omega_m}{2\pi}\int_{-\frac{\pi}{\omega_m}}^{\frac{\pi}{\omega_m}} K(t)\,e^{-ip\omega_mt} dt.
\label{e:kh}
\end{equation}
For the specific choice of $K(t)$ in Eq.~\ref{e:kdef}, we are effectively truncating the sum such that $|p| \le P=1$ and the only Fourier coefficients we obtain are $\hat{K}_0=K_0$, $\hat{K}_{+1}=\hat{K}_{-1}=dK$. From now on, we adopt the truncated notation for $p$. We also assume a harmonic solution with time-modulated amplitude and expand it in Fourier series, obtaining:
\begin{equation}
V(t)=\left(\sum_{n=-\infty}^{\infty}\hat{V}_n\,e^{i n \omega_m t}\right)e^{i \omega t},
\label{e:V}
\end{equation}
with $\omega$ being an unknown frequency at this stage, and with $\hat{V}_n$ being the Fourier coefficients of the wave amplitude. 
%
%
Differentiating $V(t)$, plugging it into Eq.~\ref{e:eom} together with $K(t)$ and simplifying $e^{i\omega t}$, yields:
\begin{equation}
\sum_{n=-\infty}^{\infty} \left[-m\left( \omega +n\omega_m \right)^2+ic\left( \omega +n\omega_m \right)\right]\hat{V}_n\,e^{in\omega_mt}+
\sum_{n=-\infty}^{\infty}\sum_{p=-P}^{P}\hat{K}_p\hat{V}_n\,e^{i (n+p) \omega_m t}=0.
\end{equation}
To simplify this expression, we pre-multiply it by $e^{ih\omega_m t}\omega_m/(2\pi)$, where $h$ is an arbitrary integer, and we integrate the result over the modulation period, from $-\pi/\omega_m$ to $\pi/\omega_m$. This averaging procedure is a standard method to study the dynamics of systems with time-varying properties, and has been adopted by others in the context of modulated media~\cite{Trainiti2016, Vila2017, Attarzadeh2018}. 
Leveraging the orthogonality of harmonic functions, we drop the summation in $n$ and obtain the following equation, valid for all values of $h$:
\begin{equation}
\left[-m\left( \omega +h\omega_m \right)^2+ic\left( \omega +h\omega_m \right)\right]\hat{V}_h+\!\!\sum_{p=-P}^{P}\hat{K}_p\hat{V}_{h-p}=0.
\label{e:eig}
\end{equation}
This system of equations needs to be solved for all integer values of $h$ to obtain an exact solution. Here, we intend to verify the validity of a truncated expansion of the solution by setting $|h| \le H=1$. Under this assumption, and recalling that $P=1$ for our choice of stiffness modulation function, Eq.~\ref{e:eig} reduces to the system of three equations:
\begin{equation}
\left(\begin{bmatrix}
\hat{K}_0 & \hat{K}_{-1} & 0\\
\hat{K}_{+1} & \hat{K}_0 & \hat{K}_{-1}\\
0 & \hat{K}_{+1} & \hat{K}_0
\end{bmatrix}-m\begin{bmatrix}
\left( \omega -\omega_m \right)^2 & 0 & 0\\
0 & \omega^2 & 0\\
0 & 0 & \left( \omega +\omega_m \right)^2
\end{bmatrix}\right.
+
\left.ic\begin{bmatrix}
\omega -\omega_m & 0 & 0\\
0 & \omega & 0\\
0 & 0 & \omega +\omega_m
\end{bmatrix}\right)
\left[\begin{matrix}
\hat{V}_{-1}\\
\hat{V}_{0}\\
\hat{V}_{+1}
\end{matrix} \right]
=
\left[ \begin{matrix}
0\\
0\\
0
\end{matrix} \right],
\label{e:eig3}
\end{equation}
which can be written in compact form as $\mathbf{D}(\omega)\,\mathbf{\hat{V}}=\mathbf{0}$. The approximated resonance frequencies of damped vibrations are the local minima of the determinant $|\mathbf{D}(\omega)|$, as shown in Fig.~\ref{f:free}(a) for parameters $dK/K_0=0.1$, $\omega_m/\omega_r=0.25$ and $\xi=0.001$.
The choice of a harmonically-modulated stiffness and a truncated solution at $|h|\le H=1$ yields three resonance frequencies for damped vibrations; these are a central frequency $\omega_r$ and two shifted ones near $\omega_r+\omega_m$ and $\omega_r-\omega_m$. 

To verify the validity of the analytical approach, we solve Eq.~\ref{e:eom} numerically using a central difference scheme, in the $0 \leq t \leq 600\,T_r$ time range, with $T_r=2\pi/\omega_r$ and time increment $dt=T_r/(10\pi)$. We choose initial conditions $[V,dV/dt]_{t=0}=[0,1]$. The normalized spectrum of the steady-state portion of the displacement signal is shown in Fig.~\ref{f:free}(b). It features a central resonance peak and multiple side-bands, as expected for modulated oscillators~\cite{Minkov2017}. One can see two main differences between the analytical and numerical results. First, the numerical results yield more peaks than the analytical approximation in Eq.\ref{e:eig3} predicted: in addition to the sidebands near $\omega_r+\omega_m$ and $\omega_r-\omega_m$, there are others near $\omega_r+2\omega_m$ and $\omega_r-2\omega_m$. Moreover, the  numerical sidebands are slightly shifted in frequency when compared to their respective eigenvalues (although this is not easy to appreciate from the figure). These inconsistencies are attributed to the 
truncation of the analytical results. 
This is discussed in more detail in Section~\ref{s:base}.

\subsection{Stability}
\label{s:stab}
When the modulations have a cosine profile in time, Eq.~\ref{e:eom} is known as Mathieu's equation. It is well known that some combinations of parameters can lead to instabilities in systems governed by this equation~\cite{Kovacic2018}. Here, we determine the regions of the modulation parameter space for which the motion of the resonator remains stable. First, we select a range of variables of interest: $0.01 \leq \omega_m/\omega_r \leq 1$ and $0 \leq dK/K_0 \leq 1$. For each $\omega_m/\omega_r$ and $dK/K_0$ couple, we solve Mathieu's equation, obtained from Eq.~\ref{e:eom} via a change of variables:
\begin{equation}
\frac{d^2V}{d\tau^2}+\bar{c}\frac{dV}{d\tau}+\left( \delta+\epsilon \cos{\tau} \right)\,V=0,
\label{e:Mat}
\end{equation}
where, for our specific problem:
\begin{equation}
\tau=\omega_mt,\,\,\,\bar{c}=2\xi\frac{\omega_r}{\omega_m},\,\,\,\delta=\frac{\omega_r^2}{\omega_m^2},\,\,\,\epsilon=2\frac{dK}{K_0}\frac{\omega_r^2}{\omega_m^2}.
\end{equation}
%
Eq.~\ref{e:Mat} is solved numerically for $\tau \in [0,2\pi]$, for two sets of initial conditions: (i) $[V,dV/d\tau]_{\tau=0}=[1,0]$, which yields displacement $V_1(\tau)$; (ii) $[V,dV/d\tau]_{\tau=0}=[0,1]$, which yields displacement $V_2(\tau)$. For each pair of $\omega_m/\omega_r$ and $dK/K_0$, according to Ref.~\cite{Kovacic2018}, the system is stable if: 
\begin{equation}
    \left|
    \mathrm{Tr}
    \begin{bmatrix}
    V_1(\tau) & V_2(\tau)\\
    dV_1(\tau)/d\tau & dV_2(\tau)/d\tau
    \end{bmatrix}_{\tau=2\pi}
    \right| < 2,
\end{equation}
where $\mathrm{Tr}$ is the trace operator.
The stability diagram as a function of the modulation frequency ratio $\omega_m/\omega_r$ and the modulation amplitude ratio $dK/K_0$ is illustrated in Fig.~\ref{f:free}(c). The shaded regions between the tongues represent the unstable regions for the damping of choice, $\xi=0.001$. One can see that the parameters used in Fig.~\ref{f:free}(a,b), corresponding to the red star-like marker in Fig.~\ref{f:free}(c), yield a stable response. The contours of the unstable regions are strongly dependent on damping. Increasing damping shrinks the unstable regions, while decreasing damping expands them. When damping is 0, the unstable tongues can extend to $dK/K_0=0$; however, one can appreciate that even an extremely small damping can guarantee stability for a wide range of parameters. This stability diagram represents an important tool to properly choose the modulation parameters.

\subsection{Base excitation}
\label{s:base}
To bridge the gap between single resonator dynamics and surface wave-metasurface interactions, we incorporate the harmonic motion of the substrate into our model. In fact, a resonator on a semi-infinite medium subject to Rayleigh waves exchanges stresses with the substrate, and these stresses are a function of the relative displacement between the base and resonator~\cite{Garova1999,Boechler2013}. At this stage, we ignore the interaction between the resonators through the substrate, and focus on the response of a single modulated oscillator to a base excitation. This is equivalent to assuming the substrate as rigid; we will consider the full problem in Section~\ref{s:saw}.

\begin{figure*}[!htb]
\centering
\includegraphics[scale=1.0]{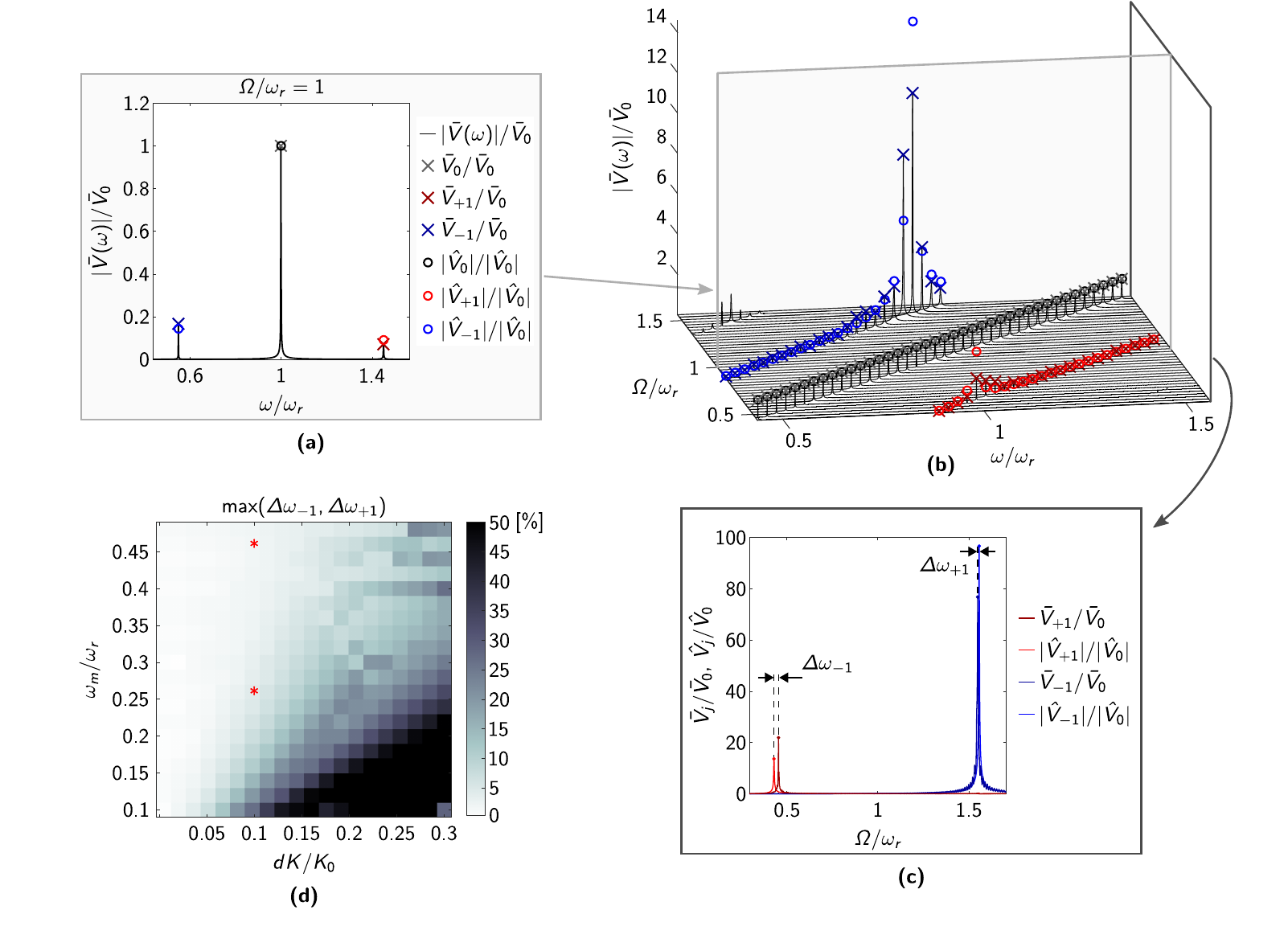}
\caption{Base excitation response of a single resonator with time-modulated stiffness. (a) Normalized Fourier transform of the numerical response of a system with $dK/K_0=0.1$, $\omega_m/\omega_r=0.45$ and $\xi=0.001$ to a harmonic base excitation of frequency $\Omega=\omega_r$. The cross markers indicate the peaks of the response, and the circular markers indicate the relative amplitudes of the Fourier coefficients. (b) Response to various base frequencies $\Omega$, where we track the numerical maxima $\bar{V}_j/\bar{V}_0$ and the relative Fourier coefficients $\hat{V}_j/\hat{V}_0$. Note that (a) is a slice of (b), and that the same legend applies to (a) and (b). (c) Evolution of the maxima of the numerical responses, and of the relative Fourier coefficients, as a function of $\Omega$. From (c), we extract the discrepancy between analytical and numerical results in predicting the frequency location of the side peaks. (d) Frequency discrepancy map.
The star markers indicate modulation parameters of interest.
}
\label{f:base}
\end{figure*}

The base excitation problem can be analyzed similarly to the free vibrations case. {Here, the forced equation of motion is a non-homogeneous version of Eq.~\ref{e:eom} and it} reads:
\begin{equation}
m\ddot{V}+c\dot{V}+K(t)V=c\dot{v}+K(t)v,
\label{e:eomb}
\end{equation}
where $v(t)=v_0\,e^{i\Omega t}$ is the harmonic base displacement, $\Omega$ the corresponding frequency of excitation and the overdot indicates a time derivative. Following the same steps detailed in Section~\ref{s:sfree} 
leads to the following system of equations: 
\begin{equation}
\left(\begin{bmatrix}
\hat{K}_0 & \hat{K}_{-1} & 0\\
\hat{K}_{+1} & \hat{K}_0 & \hat{K}_{-1}\\
0 & \hat{K}_{+1} & \hat{K}_0
\end{bmatrix}-m\begin{bmatrix}
\left( \Omega -\omega_m \right)^2 & 0 & 0\\
0 & \Omega^2 & 0\\
0 & 0 & \left( \Omega +\omega_m \right)^2
\end{bmatrix}\right.
+
\left.ic\begin{bmatrix}
\Omega -\omega_m & 0 & 0\\
0 & \Omega & 0\\
0 & 0 & \Omega +\omega_m
\end{bmatrix}\right)
\left[ \begin{matrix}
\hat{V}_{-1}\\
\hat{V}_{0}\\
\hat{V}_{+1}
\end{matrix} \right]
=
\left[ \begin{matrix}
\hat{K}_{-1}v_0\\
(\hat{K}_{0}+ic\,\Omega)v_0\\
\hat{K}_{+1}v_0
\end{matrix} \right],
\label{e:eig3b}
\end{equation}
which can be written in a compact form as $\mathbf{D}(\Omega)\,\mathbf{\hat{V}}=\mathbf{F}_b$. This expression can be solved to find three Fourier coefficients $\hat{V}_j$ for any excitation frequency $\Omega$. Coefficient $\hat{V}_0$ corresponds to frequency $\Omega$, $\hat{V}_{-1}$ to $\Omega-\omega_m$, and $\hat{V}_{+1}$ to $\Omega+\omega_m$. To quantify the accuracy of this analytical solution, we solve Eq.~\ref{e:eomb} using the same numerical procedure used in Sec.~\ref{s:sfree}. This process is illustrated in Fig.~\ref{f:base} and explained in the following.

First, we compute the numerical response to a base excitation of frequency $\Omega$. The Fourier transform of the steady-state part of the response to an excitation at $\Omega/\omega_r=1$ is shown as a continuous line in Fig.~\ref{f:base}(a), for parameters $dK/K_0=0.1$, $\omega_m/\omega_r=0.45$ and $\xi=0.001$. According to Fig.~\ref{f:free}(c), the free response of the resonator is stable for this choice of parameters. This frequency response shows several peaks: one at $\Omega/\omega_r$, and side-bands at $(\Omega+\omega_m)/\omega_r$ and $(\Omega-\omega_m)/\omega_r$. Other side-bands are also present in the numerical solution, but they are not captured by the analytical solution in Eq.~\ref{e:eig3b}. The response is normalized by the amplitude of the peak at $\Omega/\omega_r$. The peaks of interest are highlighted with cross markers in Fig.~\ref{f:base}(a). Then, we plot the analytically-derived Fourier coefficients $\hat{V}_0$, $\hat{V}_{-1}$, $\hat{V}_{+1}$, normalized by $\hat{V}_0$, at their corresponding frequencies. These are indicated as circular markers. 
We compute the response of the resonator for other values of $\Omega/\omega_r$, as presented in the waterfall plot in Fig.~\ref{f:base}(b). To quantify the discrepancy between the numerical and analytical evaluation of the frequency location of the side-bands, we track the maxima of the numerical response (cross markers) and the Fourier coefficients (circles), as a function of $\Omega/\omega_r$. This is shown in Fig.~\ref{f:base}(c), from which we calculate the discrepancy in frequency as $\max(\Delta\omega_{-1}, \Delta\omega_{+1})$, where $\Delta\omega_{-1}$ and $\Delta\omega_{+1}$ are the discrepancies between the two sets of peaks. 

This procedure is repeated for all modulation parameters of interest. We restrict our analysis to $0 \leq dK/K_0 \leq 0.3$ and $0.1 \leq \omega_m/\omega_r \leq 0.5$, all within the stable region for $\xi=0.001$. As a result, we obtain the discrepancy map of Fig.~\ref{f:base}(d). This map can be used to evaluate the error introduced by the truncated expansion of the analytical solution. It shows that there are wide parameter regions where the truncated expansion is accurate, with frequency discrepancies below 5\%. 
In light of these results, we choose the following parameters to perform the surface wave-metasurface interaction analysis: (i) $dK/K_0=0.1$ and $\omega_m/\omega_r=0.25$, which yield a discrepancy in frequency of 5\%; (ii) $dK/K_0=0.1$ and $\omega_m/\omega_r=0.45$, which yield a discrepancy in frequency of 2\%. Both sets of parameters correspond to resonators with stable responses.

\section{Surface wave dispersion in modulated metasurfaces}
\label{s:saw}

Now that we have studied the dynamics of a single resonator and learned about the acceptable ranges for the modulation parameters, we tackle the problem of a spatio-temporally modulated metasurface. Here, we couple the motion of an elastic substrate with the array of modulated resonators using an effective medium approach~\cite{Garova1999} and a truncated plane-wave expansion of the solution~\cite{Vila2017}. To quantify the dispersion characteristics of the modulated metasurface, we use a first-order asymptotic analysis~\cite{Nassar2017prsa, Nassar2017eml}. 

\subsection{Analytical dispersion relation of non-modulated metasurfaces}
\label{s:metasurf}

We begin our investigation by first recalling the dynamics of vertically polarized surface waves (of the Rayleigh type) propagating in an isotropic, elastic, homogeneous medium of infinite depth, decorated with an array of vertically-vibrating resonators. We restrict our analysis to plane waves propagating in the $x,z$ plane (see Fig.~\ref{f:met}), and we assume plane-strain conditions. The displacements along $x$ and $z$ are called $u$ and $v$, respectively. In the absence of body forces, pressure and shear waves propagating in the substrate are described by the wave equations~\cite{Graff1991}:
\begin{subequations} 
\begin{equation} \label{e:bulk 1}
    \nabla^{2} \Phi=\frac{1}{c_{L}^{2}} \frac{\partial^{2} \Phi}{\partial t^{2}},
\end{equation}
\begin{equation} \label{e:bulk 2}
    \nabla^{2} \Psi_{y}=\frac{1}{c_{S}^{2}} \frac{\partial^{2} \Psi_{y}}{\partial t^{2}},
\end{equation}
\end{subequations}
where the dilational $\Phi$  and  the transverse $\Psi_{y}$ potentials are introduced via Helmholtz decomposition of the substrate displacement field, $u=\frac{\partial\Phi}{\partial x}-\frac{\partial\Psi_y}{\partial z}$ and  $v=\frac{\partial\Phi}{\partial z}+\frac{\partial\Psi_y}{\partial x}$. The pressure ($c_{L}$) and shear ($c_{S}$) wave velocities are given as:
\begin{equation}
    c_{L}=\sqrt{\frac{\lambda+2\mu}{\rho}}, \quad c_{S}=\sqrt{\frac{\mu}{\rho}},
\end{equation}
where $\lambda$ and $\mu$ are the elastic Lam\'e constants and $\rho$ is the mass density of the substrate. Following a standard approach for the derivation of Rayleigh waves dispersion, we assume the following form of the potentials:
\begin{subequations} 
\begin{equation} \label{e:pot 1}
    \Phi=A_{0}\,e^{\sqrt{k^2-{\omega^{2}}/{c_{L}^{2}}}\,z}\,e^{i(\omega t-kx)}, 
\end{equation}
\begin{equation} \label{e:pot 2}
    \Psi_{y}=B_{0}\,e^{\sqrt{k^2-{\omega^{2}}/{c_{S}^{2}}}\,z}\,e^{i(\omega t-kx)},
\end{equation}
\label{e:pot}
\end{subequations}
with $k$ being the wavenumber along $x$.
%
%

In parallel, we account for the presence of the surface resonators. This is done by considering the equation of motion of an undamped resonator placed on the free surface (corresponding to $z=0$) and excited by the substrate motion $v(x,0,t)=v_{0}$:
\begin{equation}
m\ddot{V}+K_0(V-v_{0})=0.
\label{e:eom2}
\end{equation}
Following the procedure adopted in Ref.~\cite{Garova1999}, we assume a harmonic motion $V=V_0\,e^{i(\omega t-kx)}$ for the resonator and consider the normal stress exerted by the resonator at the surface as its inertial force divided by the footprint area $A=s^2$, where $s$ is the distance between resonators, i.e., the unit cell size of the array. This stress is defined as:
\begin{equation} \label{e:average stress}
    \sigma_{zz,r}=-\frac{m}{A}\ddot{V}=\frac{m}{A}\omega^2 V.
\end{equation}
By using this assumption, often referred to as effective medium approach~\cite{Boechler2013}, we restrict our analysis to wave propagation regimes where the surface wavelengths are much larger than the characteristic resonator spacing, $s$. The average stress in Eq.~\eqref{e:average stress} can be used as a boundary condition for the normal stress of the elastic half-space at $z=0$:
\begin{subequations} 
\begin{equation} \label{e:normal stress bc at z=0}
    \sigma_{zz}=\sigma_{zz,r},
\end{equation}
together with the free stress condition on the tangential component: 
\begin{equation} \label{e:tang. stress bc at z=0}
    \sigma_{zx}=0.
\end{equation}
\end{subequations}
For a linear elastic and isotropic material, the stresses can be related to the potentials $\Phi$ and $\Psi_y$ using the constitutive relations~\cite{Graff1991}:
\begin{subequations}
	\begin{align}
		\label{sigzx}
			\sigma_{zx} &= \mu \left(2\frac{\partial^2\Phi}{\partial x \partial z} + \frac{\partial^2\Psi_y}{\partial x^2 } - \frac{\partial^2\Psi_y}{\partial z^2 }\right),
		\\
		\label{sigzz}
			\sigma_{zz} &= (\lambda+2\mu) \left(\frac{\partial^2\Phi}{\partial z^2 }+ \frac{\partial^2\Psi_y}{\partial x \partial z}\right) + \lambda \left(\frac{\partial^2\Phi}{\partial x^2 } - \frac{\partial^2\Psi_y}{\partial x \partial z}\right).
	\end{align}
	\label{e:sig}
\end{subequations}

At this stage, using Eq.~\eqref{e:sig}, we express the boundary conditions  in Eq.~\eqref{e:normal stress bc at z=0} and Eq.~\eqref{e:tang. stress bc at z=0} in terms of surface wave potentials in Eqs.~\eqref{e:pot}, and obtain the expressions:
\begin{subequations}
\begin{equation}
\left[-2i\mu \, \sqrt{k^{2}-\frac{\omega^{2}}{c_{L}^{2}}}\,k A_0 + \mu\left(\frac{\omega^2}{c_S^2} - 2k^2\right) B_0\right]\,e^{i(\omega t-kx)}=0,
\end{equation}
\begin{equation}
\left[\left(2\mu k^{2}-2\mu\frac{\omega^{2}}{c_{L}^{2}} - \lambda\frac{\omega^2}{c_{L}^2}\right)A_0 -2i\mu k \sqrt{k^{2}-\frac{\omega^{2}}{c_{S}^{2}}}\,B_0 -m\frac{\omega^2}{A}V_0\right]\,e^{i(\omega t-kx)}=0.
\end{equation}
\end{subequations}
Coupling these two equations with the equation of motion of the resonator, Eq.~\eqref{e:eom2}, and dropping the exponential $e^{i(\omega t-kx)}$, we obtain:
\begin{equation}
\label{e:metasurf}
\left[\begin{array}{ccc}
{-2i\mu k\sqrt{k^{2}-\frac{\omega^{2}}{c_{L}^{2}}}} & {\mu(\frac{\omega^2}{c_S^2} - 2k^2)} & {0} \\
{2\mu (k^{2}-\frac{\omega^{2}}{c_{L}^{2}}) - \lambda\frac{\omega^2}{c_{L}^2}} & {-2i\mu k \sqrt{k^{2}-\frac{\omega^{2}}{c_{S}^{2}}} } & {-m\frac{\omega^2}{A}} \\
{-K_{0}\sqrt{k^{2}-\frac{\omega^{2}}{c_{L}^{2}}}} & {i K_0 k} &{-m\omega^2 + K_0}
\end{array}\right]\left[\begin{array}{ccc}{A_0}\\{B_0}\\{V_0}\end{array}\right]=
\left[\begin{array}{ccc}{0}\\{0}\\{0}\end{array}\right].
\end{equation}
This system of three equations can be written in compact form as $\boldsymbol{\Pi}(k,\omega)\,\mathbf{q}_0=\mathbf{0}$. It represents the necessary condition for the plane-wave solutions to hold.
Non-trivial solutions of Eq.~\ref{e:metasurf} are found by setting $|\boldsymbol{\Pi}(k,\omega)|=0$, which yields the non-modulated metasurface dispersion relation. An example of this dispersion relation is given by the solid black lines in Fig.~\ref{f:disp}(a), for an elastic substrate with $c_L/c_S=1.5$ and a metasurface with mass ratio $m \omega_r/(A \rho c_S)=0.15$. 
Note that the coupling between Rayleigh waves and surface resonators induces a subwavelenght bandgap in the surface wave spectrum. This gap covers the frequency range $\omega_r < \omega < \omega_r(\beta+\sqrt{\beta^2+1})$, where $\beta=\frac{m\omega_r}{2\rho A c_S}\sqrt{1-c_S^2/c_L^2}$ ~\cite{Palermo2016}. Further details about the dispersive features of a non-modulated metasurface can be found in Refs.~\cite{Garova1999, Boechler2013, Palermo2016}.

\subsection{Analytical dispersion relation of modulated metasurfaces}
\label{s:modmetasurf}

We consider a plane-wave spatio-temporal modulation of the stiffness of the resonators:
\begin{equation}
K(t)=K_0+2dK \cos{\left( \omega_m t -k_m x \right)},
\label{e:kdef meta}
\end{equation}
where $k_m$ is the modulation wavenumber. The key difference between Eqs.~\ref{e:kdef} and~\ref{e:kdef meta} is the presence of a spatially-varying phase term, $k_mx$. {Note that such one-dimensional modulation restricts our investigation to those scenarios where the surface wave is collinear with the direction of the stiffness modulation.}
This spatial modulation of the stiffness parameter, on its own, results in the appearance of a symmetric frequency gap in the dispersion relation of the surface waves (symmetric with respect to $k$). When combined with the temporal modulations, these frequency gaps occur at different frequencies for forward- and backward-traveling waves, i.e., non-reciprocal propagation emerges~\cite{Swinteck2015,Trainiti2016,nassarPRB}.

Based on the results of Section~\ref{s:sdof}, we choose a modulation amplitude $dK/K_0$ and frequency $\omega_m/\omega_r$ such that the response of the resonators remain stable, and the truncated approximation of the response is acceptable. 
To ensure stability in the presence of spatio-temporal modulations, we need to additionally check that the modulation wave speed is smaller than the phase velocity of the medium~\cite{Cassedy1963}, i.e., $\omega_m/k_m<c\,(\omega)$. This condition might not be respected near $\omega_r$ if the resonant bandgap is at very low frequencies. Note, however, that our results on the stability of a single resonator in Fig.~\ref{f:free}(c) already warned us to stay away from values of the modulating frequency that are close to $\omega_r$.

The modulating wave generates a scattered wavefield, here described by the vector of amplitudes $\mathbf{q}_j=[\hat{A}_j, \hat{B}_j, \hat{V}_j]^T$, where $j$ is a non-zero integer. These amplitudes are associated to the substrate potentials:
\begin{subequations} \label{e:potential function j}
\begin{equation}
    \Phi_j=\hat{A}_{j}\,e^{\sqrt{k_j^{2}-{\omega_j^{2}}/{c_{L}^{2}}}\,z}\,e^{i(\omega_j t-k_j x)}, 
\end{equation}
\begin{equation}
    \Psi_{y,j}=\hat{B}_{j}\,e^{\sqrt{k_j^{2}-{\omega_j^{2}}/{c_{S}^{2}}}\,z}\,e^{i(\omega_j t-k_j x)},
\end{equation}
\end{subequations}
and to the resonator displacement:
\begin{equation}
   V_{j}=\hat{V}_{j}\,e^{i(\omega_j t-k_j x)},
\end{equation}
For convenience, we define the shifted frequency and wavenumber as:
\begin{equation}
\omega_j=\omega+j\omega_m,  \quad k_j=k+j k_m. 
\end{equation}
%
%

\begin{figure*}[!htb]
\centering
\includegraphics[scale=1]{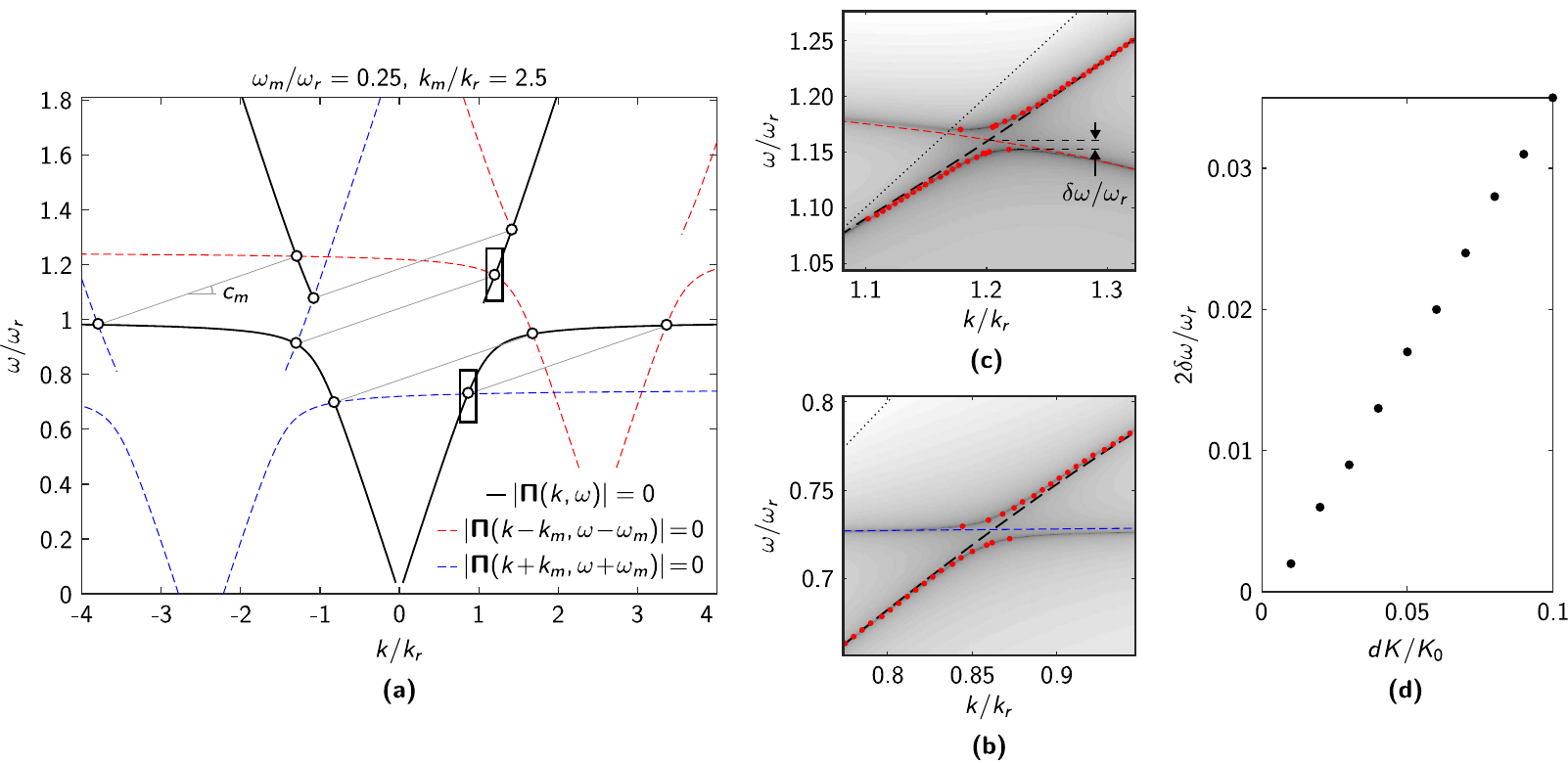}
\caption{Dispersion properties of modulated and non-modulated metasurfaces. (a) Dispersion curves. The solid black curves represent the non-modulated dispersion relation, while the dashed red and blue lines are the shifted curves for $j=-1$ and $j=+1$, respectively, for modulation parameters $\omega_m/\omega_r=0.25$ and $k_m/k_r=2.5$. The crossing points are highlighted with circular markers. The thin gray lines connect phase-matched points of the original dispersion curves. (b), (c) Details of the crossing points that are highlighted by boxes in (a). The dark regions of the colormap follow the minima of the determinant of Eq.~\eqref{e:metasurf mod}, while the circular red markers indicate the asymptotic evaluation of the modulated dispersion. The thin dotted line represents the sound cone. All cases correspond to modulation amplitude $dK/K_0=0.05$. (b) A case of veering, where no frequency band gap is found.  (c) A case of locking that features a frequency bandgap of normalized width $2\delta \omega/\omega_r$. (d) Evolution of the width of the bandgap in (c) as a function of the modulation amplitude.}
\label{f:disp}
\end{figure*}

The scattered field has a non-negligible amplitude only when the phase matching condition  $|\boldsymbol{\Pi}(k,\omega)|$=$|\boldsymbol{\Pi}(k_j,\omega_j)|$=0 is met~\cite{Nassar2017prsa}, namely at the crossing points between the original dispersion curves $|\boldsymbol{\Pi}(k,\omega)|$=0 and the shifted curves $|\boldsymbol{\Pi}(k+j k_m,\omega+j \omega_m)|=0$. A graphical representation of two shifted curves for $j=\pm 1$ is provided in Fig.~\ref{f:disp}(a) for a metasurface modulated with frequency $\omega_m/\omega_r=0.25$ and wavenumber $k_m/k_r=2.5$, where $k_r=\omega_r/c_S$. 
The asymmetric positioning of the crossing points between regions with positive and negative wavenumbers suggests the occurrence of direction-dependent phenomena within the metasurface. We predict the dispersion properties of the modulated meatsurface near these crossing points using a truncated plane-wave expansion. In particular, we assume that the surface wave potentials have the following form, comprising non-modulated and scattered amplitudes:
\begin{subequations} 
\begin{equation} \label{e:pot 1 PW}
    \Phi=\hat{A}_{0}\,e^{\sqrt{k^2-{\omega^{2}}/{c_{L}^{2}}}\,z} \,e^{i(\omega t-k x)} + \sum^{1}_{\substack{j=-1 \\ j \neq 0}}\hat{A}_{j}\,e^{\sqrt{k_j^{2}-{\omega_j^{2}}/{c_{L}^{2}}}\,z} \,e^{i(\omega_j t-k_j x)},
\end{equation}
\begin{equation} \label{e:pot 2 PW}
    \Psi_y=\hat{B}_{0}\,e^{\sqrt{k^2-{\omega^{2}}/{c_{S}^{2}}}\,z} \,e^{i(\omega t-k x)} + \sum^{1}_{\substack{j=-1 \\ j \neq 0}}\hat{B}_{j}\,e^{\sqrt{k_j^{2}-{\omega_j^{2}}/{c_{S}^{2}}}\,z} \,e^{i(\omega_j t-k_j x)}, 
\end{equation}
\end{subequations}
and a resonator displacement:
\begin{equation} \label{e: res PW}
   V=\hat{V}_{0}\,e^{i(\omega t-k x)} + \sum^{1}_{\substack{j=-1 \\ j \neq 0}}\hat{V}_{j}\,e^{i(\omega_j t-k_j x)}.
\end{equation}
The choice of $j=\pm1$ is direct consequence of using a harmonic plane-wave modulation in Eq.~(\ref{e:kdef meta}); otherwise, higher-order terms need to be included. 

Following the same procedure adopted for the
non-modulated case, we substitute the expanded potentials, Eq.~\eqref{e:pot 1 PW} and Eq.~\eqref{e:pot 2 PW}, into the constitutive equations, Eq.~\eqref{sigzx} and Eq.~\eqref{sigzz}. Similarly, we use the truncated resonator displacement, Eq.~\eqref{e: res PW}, in the governing equation of the resonator, Eq.~\eqref{e:eom2}, and boundary condition, Eq~\eqref{e:average stress}. The result is finally substituted into the boundary conditions, Eq.~\eqref{e:normal stress bc at z=0} and Eq.~\eqref{e:tang. stress bc at z=0}. After collecting and simplifying the common exponential in each equation, we obtain:
\begin{equation}
\label{e:metasurf mod}
\left[\begin{array}{ccc}
{\boldsymbol{\Pi}(k_{-1},\omega_{-1})}&{\boldsymbol{\Gamma}(k,\omega)} &\mathbf{0}\\
{\boldsymbol{\Gamma}(k_{-1},\omega_ {-1})}&{\boldsymbol{\Pi}(k,\omega)}&{\boldsymbol{\Gamma}(k_{+1},\omega_{+1})}\\
\mathbf{0}&{\boldsymbol{\Gamma}(k,\omega)} &{\boldsymbol{\Pi}(k_{+1},\omega_{+1})}
\end{array}\right]\left[\begin{array}{ccc}{\mathbf{q}_{-1}}\\{\mathbf{q}_0}\\ \mathbf{q}_{+1}\end{array}\right]=\mathbf{0},
\end{equation}
where the submatrix $\boldsymbol{\Pi}$ is defined in Eq.~\eqref{e:metasurf}, and the submatrix $\boldsymbol{\Gamma}$ is defined as:
%
%
%
\begin{equation} \label{e:Gamma}
\boldsymbol{\Gamma}(k,\omega)=\left[\begin{array}{ccc}
{0} & {0} & {0} \\
{0} & {0} & {0} \\
{-dK\sqrt{k^{2}-\frac{\omega^{2}}{c_{L}^{2}}}} & {i\,dK\,k} & {dK}
\end{array}\right].
\end{equation}

{\noindent Note that the operator $\boldsymbol{\Gamma}(k_j,\omega_j)$ describes the coupling between the scattered $j$ and fundamental $0$ wave fields introduced by the stiffness modulation of the resonator.}

The expression in Eq.~\eqref{e:metasurf mod},  written in compact form as $\mathbf{\Lambda}(k,\omega)\,\mathbf{q}=\mathbf{0}$, describes the relation between the Rayleigh waves and the modulation-induced scattered waves. This relation is valid when the scattered field interacts strongly with the main field, i.e., near the crossings of non-modulated and translated dispersion curves, as indicated in Fig.~\ref{f:disp}(a). 
Nontrivial solutions of Eq.~\eqref{e:metasurf mod} are obtained by setting the determinant of the $9\times9$ matrix equal to 0, $|\mathbf{\Lambda}(k,\omega)|=0$.
The resulting equation describes the dispersion relation of the modulated system in the vicinity of the crossing points between the fundamental and the shifted dispersion curves. We refrain from seeking a closed-form expression of its roots. Nevertheless, by evaluating the determinant $|\mathbf{\Lambda}(k,\omega)|$ in the neighborhood of the crossing points, and finding its local minima, we can identify the dispersion branches for the modulated system. Examples of modulated branches are provided in Fig.~\ref{f:disp}(b,c), where the magnitude of $|\mathbf{\Lambda}(k,\omega)|$ near the two crossing points is displayed as a colormap, with the minima being darker. In the neighborhood of the crossing points, the modulated branches are characterized by frequency ($\delta \omega$) and wavenumber ($\delta k$) shifts with respect to the intersection of the fundamental ($|\mathbf{\Pi}(k,\omega)|=0$) and translated ($|\boldsymbol{\Pi}(k+j k_m,\omega+j \omega_m)|=0$) dispersion curves. These shifts result from the repulsion between the two interacting modes.  

The pair ($\delta k,\delta \omega$) can be calculated as the leading-order correction to ($k,\omega$) in an asymptotic analysis of the problem ~\cite{Nassar2017prsa,Hinch}. 
For this purpose, we expand the surface wave potentials and the resonator displacement around the crossing point of interest, as shown in the following:
\begin{subequations} 
\begin{equation} \label{e:pot 1 cor}
    \tilde{\Phi}=\left(\tilde{A}_{0}\,e^{\sqrt{(k+\delta k)^2-{(\omega+\delta \omega)^{2}}/{c_{L}^{2}}}\,z} \,e^{i(\omega t-k x)}+\tilde{A}_{j} \,e^{\sqrt{(k_j+\delta k)^{2}-{(\omega_j+\delta \omega)^{2}}/{c_{L}^{2}}}\,z}\,e^{i(\omega_j t-k_j x)}\right)\,e^{i(\delta\omega t-\delta k x)}, 
\end{equation}
\begin{equation} \label{e:pot 2 cor}
    \tilde{\Psi}_y=\left(\tilde{B}_{0}\,e^{\sqrt{(k+\delta k)^2-{(\omega+\delta \omega)^{2}}/{c_{S}^{2}}}\,z} \,e^{i(\omega t-k x)}+\tilde{B}_{j} \,e^{\sqrt{(k_j+\delta k)^{2}-{(\omega_j+\delta \omega)^{2}}/{c_{S}^{2}}}\,z}\,e^{i(\omega_j t-k_j x)}\right)\,e^{i(\delta\omega t-\delta k x)}, 
\end{equation}
\begin{equation} \label{e: res cor}
   \tilde{V}=\left(\tilde{V}_{0}\,e^{i(\omega t-k x)}+\tilde{V}_{j}\,e^{i(\omega_j t-k_j x)}\right)\,e^{i(\delta\omega t-\delta k x)},
\end{equation}
\end{subequations}
where $j$ is either +1 or -1 depending on which shifted branch satisfies the phase matching condition with the fundamental dispersion curve.
With these ansatzes, and replicating the procedure we used to obtain the dispersion relation for the modulated metasurface, we obtain:
\begin{equation}
\label{e:metasurf correction}
\left[\begin{array}{ccc}
{\boldsymbol{\Pi}}(k+\delta k,\omega+\delta \omega) & {\boldsymbol{\Gamma}(k_j+\delta k,\omega_j+\delta \omega)}  \\
{{\boldsymbol{\Gamma}}(k+\delta k,\omega+\delta \omega) } & {\boldsymbol{\Pi}(k_j+\delta k,\omega_j+\delta \omega)}
\end{array}\right]\left[\begin{array}{ccc}{\mathbf{q}_0}\\ \mathbf{q}_j\end{array}\right]=\mathbf{0},
\end{equation}
%
We can then find the corrections $\delta k$ and $\delta \omega$  by setting the determinant of the $6\times6$ matrix in Eq.~\eqref{e:metasurf correction} to zero.
Further details on this computation are given in~\ref{a:analy}. 

Examples of corrected portions of the dispersion relation are shown in Fig.~\ref{f:disp}(b,c) as red dotted curves. We can see that the corrections are non-zero only in the neighborhood of the crossing points, and that they show an excellent agreement with the minima of the determinant of the matrix in Eq.~\eqref{e:metasurf mod}. 

\subsection{Physical insight on the modulated dispersion relation}

From Fig.~\ref{f:disp}(b,c), we observe that the presence of a spatio-temporal modulation causes the fundamental and shifted dispersion curves to repel each other. Two distinct phenomena are observed depending on whether the fundamental and shifted branches propagate along the same direction or not, i.e., whether the group velocities $c_g={\partial \omega}/{\partial k}$ and $c_{gj}={\partial \omega_j}/{\partial k_j}$ satisfy $c_{g}c_{gj}>0$ or $c_{g}c_{gj}<0$, respectively. For a pair of co-directional branches like those shown in  Fig.~\ref{f:disp}(b),  the interacting modes veer without crossing as a result of the repulsion between the fundamental and scattered modes. No significant frequency shift is found and consequently no directional band gaps are generated.
Conversely, for a couple of contra-directional branches, as shown in Fig.~\ref{f:disp}(c), the repulsion between the pair of coupled modes results in a branch locking phenomenon~\cite{mace2012} and, in some occasions, in the opening of a directional bandgap. We quantify the branch repulsion by evaluating the bandgap width at the locking point, $2\delta \omega$, as a function of the modulation amplitude, $dK$. As expected from the first-order nature of the correction terms in Section~\ref{s:modmetasurf}, the width of a directional bandgap is proportional to the modulation amplitude; see Fig.~\ref{f:disp}(d). 

We remark that for any crossing point $(k^*,\,\omega^*)$ at the intersection of $|\boldsymbol{\Pi}(k,\omega)|=0$ and $|\boldsymbol{\Pi}(k+ k_m,\omega+\omega_m)|=0$, we can identify a crossing point $(\omega^*+\omega_m,\,k^*+k_m)$, e.g., at the intersection of $|\boldsymbol{\Pi}(k,\omega)|=0$ and $|\boldsymbol{\Pi}(k- k_m,\omega- \omega_m)|=0$, that is phase-matched to $(k^*,\,\omega^*)$ via the pumping wave~\cite{Nassar2017eml}. In Fig.~\ref{f:disp}(a), all crossing points connected by thin gray lines are phase-matched, being only separated by a $\pm(k_m,\,\omega_m)$ translation. According to Eq.~\eqref{e:pot 1 cor} and Eq.~\eqref{e:pot 2 cor} we expect that for a surface wave traveling within the modulated metasurface with frequency $\omega^*$ and wavenumber $k^*$, a scattered field is generated with modulated frequencies and wavenumber $(\omega^*+\omega_m,\,k^*+k_m)$. 
Similarly, for a fundamental surface wave at $(\omega^*+\omega_m,\,k^*+k_m)$, a scattered field at $(\omega^*,\,k^*)$ is expected.
In other words, if we send a wave at a frequency near one of the crossings, the metasurface will generate waves at the frequency of the corresponding phase-matched point~\cite{Nassar2017prsa}. Numerical evidence of this intriguing dynamic behavior, that hints to the possibility of using modulated metasurfaces as frequency converters for surface waves, is provided in Section~\ref{s:nr}.

\section{Surface wave non-reciprocity and other modulation-induced effects}
\label{s:nr}
\begin{figure*}[!htb]
\centering
\includegraphics[scale=1]{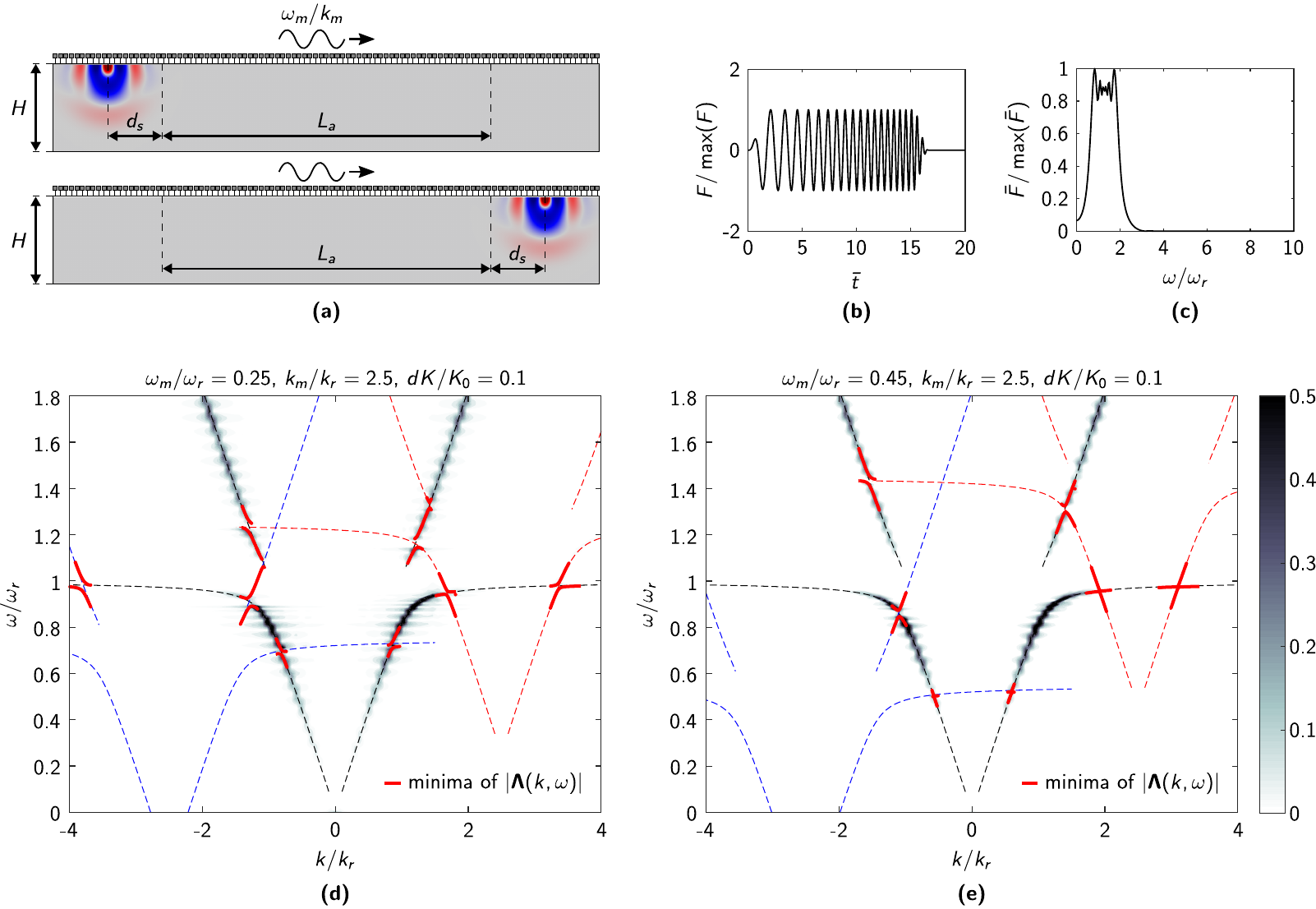}
\caption{Numerical reconstruction of the modulated dispersion curves. (a) Schematic of the numerical models for right-going and left-going surface waves, with a right-going modulating wave. (b) Time history and (c) frequency content of the point force applied at the source. (d) Dispersion curves reconstructed via a 2D-DFT of the space-time evolution of the vertical displacement on the surface underneath the resonators, $v(x,0,t)$. The system has modulation parameters $dK/K_0=0.1$, $\omega_m/\omega_r=0.25$ and $k_m/k_r=2.5$. The colormap is scaled with respect to its maximum value. The analytical dispersion, shown as a thick red line, is obtained by tracing the local minima of $|\mathbf{\Lambda}(k,\omega)|$ in a range $\pm 0.1 k$ and $\pm 0.1 \omega$ around each crossing point. The dispersion curves of the non-modulated metasurface, $|\boldsymbol{\Pi}(k,\omega)|=0$, and its shifted twins, $|\boldsymbol{\Pi}(k+k_m,\omega+\omega_m)|=0$ and $|\boldsymbol{\Pi}(k-k_m,\omega-\omega_m)|=0$, are shown as black, red and blue dashed lines, respectively. (e) Same as (d), for modulation parameters $dK/K_0=0.1$, $\omega_m/\omega_r=0.45$ and $k_m/k_r=2.5$.}
\label{f:disp_num}
\end{figure*}

We now resort to finite element (FE) simulations to analyze the propagation of surface waves in a modulated metasurface and to validate the directional behavior predicted by our analytical model. Our 2D plane-strain FE model, implemented in COMSOL Multiphysics, consists of a portion of an elastic substrate of depth $H=4\lambda_0$, where $\lambda_0=\omega_r/c_R$ and $c_R$ is the Rayleigh wave velocity in the substrate. One of our models is sketched in Fig.~\ref{f:disp_num}(a). 
The substrate features an array of resonators mounted on its free surface with spacing $s=\lambda_0/23$. All edges of the domain, apart from the one decorated with resonators, are characterized by low-reflecting boundary conditions. A convergent mesh of quadratic Lagrangian elements is used to discretize the substrate and to ensure that the wave field is accurately captured in the frequency range of interest. The stiffness of each resonator varies in space and time according to Eq.~\eqref{e:kdef meta}. Based on the previous considerations on accuracy and stability in Section~\ref{s:sdof}, we choose modulation parameters $dK=0.1\,K_0$, $k_m=2.5\,k_r$ and either $\omega_m=0.25\,\omega_r$ or $\omega_m=0.45\,\omega_r$.  

\subsection{Numerical dispersion reconstruction}

We perform transient simulations to numerically reconstruct the dispersion properties of the modulated metasurface, using the models shown in Fig.~\ref{f:disp_num}(a). We excite the medium with a vertical sine-sweep point force having frequency content $0.5\,\omega_r<\omega<2\,\omega_r$, as shown in Fig.~\ref{f:disp_num}(b,c). We record the vertical surface displacement $v(x,0,t)$ at 1000 equally-spaced locations along a length $L_a=15\lambda_0$ for a normalized time $0 < \bar{t} < 125$, where $\bar{t}=t/T_r$ and $T_r=2\pi/\omega_r$. To reconstruct the dispersion branches for $k>0$ and $k<0$, we simulate both a right-propagating (top panel of Fig.~\ref{f:disp_num}(a)) and a left-propagating wave (bottom panel), with a modulating wave that is always right-propagating. In both cases, the source is placed at a distance $d_s=5\lambda_0$ from the closest recording point. The recorded space-time traces are then transformed via 2D Discrete Fourier Transform (2D-DFT) to obtain the wavenumber-frequency spectrum $\bar{v}(k,0,\omega)$. By following the higher-amplitude regions of this two-dimensional spectrum, we can identify the numerical dispersion branches.

The reconstructed dispersion for modulation parameters $dK=0.1\,K_0$, $\omega_m=0.25\,\omega_r$ and $k_m=2.5\,k_r$ is shown as a colormap in Fig.~\ref{f:disp_num}(d). The analytical dispersion, shown as a thick red line, is obtained by tracing the minima of $|\mathbf{\Lambda}(k,\omega)|$ near the crossing points. For convenience, we also replicate on the same figure the original (non-modulated) dispersion curve and its shifted analogs (thin dashed lines). This plot unequivocally illustrates that the dispersive features observed in the numerical results are consistent with the analytical predictions. In particular, one can see that the numerical results clearly indicate the presence of several modulation-induced features: (i) two coupled directional bandgaps of narrow extent at $0.69\,\omega_r$ for left-propagating and $0.93\,\omega_r$ for right-propagating waves; (ii) two coupled veering points at $0.73\,\omega_r$ and $0.98\,\omega_r$, both for right-propagating waves; (iii) two coupled and relatively-wide directional gaps at $0.92\,\omega_r$ and $1.17\,\omega_r$ for left- and right-propagating waves, respectively. 
We repeat this reconstruction procedure for different modulation parameters: $dK=0.1\,K_0$, $\omega_m=0.45\,\omega_r$ and $k_m=2.5\,k_r$. The results are shown in Fig.~\ref{f:disp_num}(e), and they display a similar consistency with the analytical predictions as for the previous configuration. In this case, the features of interest are two coupled directional gaps at the locking frequencies $0.86\,\omega_r$ and $1.31\,\omega_r$, for left- and right-propagating waves, respectively. These gaps are of interest because they are characterized by a significant reduction in spectral amplitude.

\subsection{Non-reciprocal transmission and conversion-by-reflection}
\label{s:nrtr}

\begin{figure*}[!htb]
\centering
\includegraphics[scale=1]{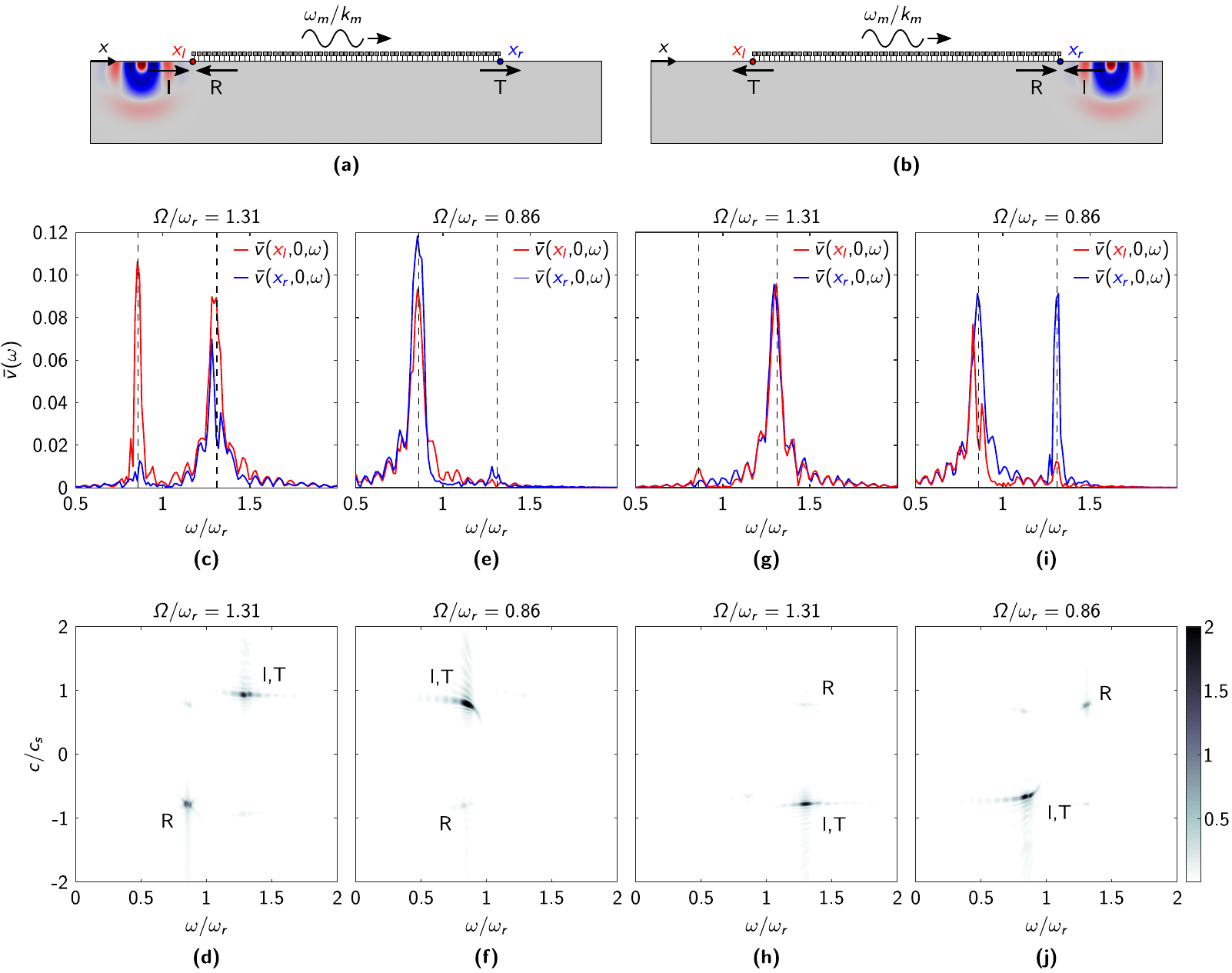}
\caption{Transient FE simulations of the propagation of narrow-band signals centred at the directional gap frequencies ($0.86\,\omega_r$ and $1.31\,\omega_r$, Fig.~\ref{f:disp_num}(e)) through a modulated metasurface.  Schematic of the numerical setup for (a) right-propagating and (b) left-propagating surface waves. Spectral content of the vertical surface wave field recorded at the left and right edges of the resonators array for: (c) right-propagating waves at $\Omega=1.31\,\omega_r$, (e) right-propagating waves at $\Omega=0.86\,\omega_r$, (g) left-propagating waves at $\Omega=1.31\,\omega_r$, (i) left-propagating waves at $\Omega=0.86\,\omega_r$. Radon transform of time-space surface wave records computed along the resonator array for: (d) right-propagating waves at $\Omega=1.31\,\omega_r$, (f) right-propagating waves at $\Omega=0.86\,\omega_r$, (h) left-propagating waves at $\Omega=1.31\,\omega_r$, (j) left-propagating waves at $\Omega=0.86\,\omega_r$.}
\label{f:TR}
\end{figure*}

To verify the characteristics of the scattered field responsible for directional wave propagation, we perform transient simulations with narrow-band waveforms centered at those frequencies. For these analyses, we use the models shown in Fig. \ref{f:TR}(a,b), cf. Fig.~\ref{f:disp_num}(a). In both cases, we have two substrate-only regions separated by a region of length $L_a=12.5\,\lambda_0$ that features a large number of surface resonators (286) spaced at $s=\lambda/23$. The response is recorded at locations $x_l$ and $x_r$, that mark the left and right edges of the region with resonators, respectively. In both configurations, the point source is located on the free surface at a distance $d_s=3.5\,\lambda$ from the corresponding edge of the resonators region. In all cases, the modulating wave is right-propagating, with $dK=0.1\,K_0$, $\omega_m=0.45\,\omega_r$ and $k_m=2.5\,k_r$. This corresponds to the dispersion curve in Fig.~\ref{f:disp_num}(e). 

We begin our investigation by considering a right-propagating surface wave (i.e., incident to the array at $x_l$) at frequency $\Omega=1.31\,\omega_r$. The spectra of the time signals recorded at $x_l$ and $x_r$ are shown in Fig.~\ref{f:TR}(c). The spectrum at $x_r$ (blue line), corresponding to a wave transmitted through the array of resonators, shows a significant amplitude reduction at $\Omega=1.31\,\omega_r$, in agreement with the directional gap predicted by our analysis. The amplitude gap is accompanied by the generation of a side peak at the twin locking frequency $0.86\,\omega_r$. This frequency content appears even more markedly in the spectrum of the signal recorded at the $x_l$ location (red line). This second peak corresponds to the reflected field caused by the modulated array of resonators. To support this claim, we compute the two-dimensional Radon transform (wave speed $c$ versus frequency $\omega$) of the time-space data matrix recorded within the array of resonators. By means of this transform, we determine if a signal with a certain frequency content is right-propagating (positive $c$) or left-propagating (negative $c$). The amplitude of this spectrum, shown as a colormap in Fig.~\ref{f:TR}(d), confirms that the signal content at $0.86\,\omega_r$ travels from right to left, opposite to the direction of the incident signal at $1.31\,\omega_r$. This indicates that the modulated metasurface can convert an incident wave into a reflected wave with a different frequency content---shifted from the original frequency by the modulating one~\cite{Nassar2017eml}. To verify non-reciprocity, we send a left-propagating wave with frequency centered at $1.31\,\omega_r$. In this case, the signal travels undisturbed through the metasurface, as confirmed by the spectra at $x_l$ and $x_r$, shown in Fig.~\ref{f:TR}(g). Moreover, no evidence of reflected waves is found in the Radon transform shown in Fig.~\ref{f:TR}(h).

We replicate these analyses for left- and right-propagating surface waves excited at the phase-matched locking frequency $\Omega=0.86\,\omega_r$. In this case, left-propagating waves travel almost undisturbed within the metasurface, as confirmed by the spectral contents in Fig.~\ref{f:TR}(e) that feature waves at the carrier frequency only, and by the Radon transform in Fig.~\ref{f:TR}(f). Conversely, the directional gap for right-propagating waves causes an attenuation of the transmitted signal at $0.86\,\omega_r$, as shown by the red line of Fig.~\ref{f:TR}(i). This phenomenon is accompanied by a back-scattering of the coupled frequency $1.31\,\omega_r$, as indicated by the blue line in Fig.~\ref{f:TR}(i) and by the Radon transform in Fig.~\ref{f:TR}(j).

While this section has been dedicated to the response to excitation frequencies within the directional bandgaps,  the reader can find details on the response of a metasurface excited at a veering point in~\ref{a:transm}.

\subsection{Surface-bulk wave conversion}

\begin{figure*}[!htb]
\centering
\includegraphics[scale=1]{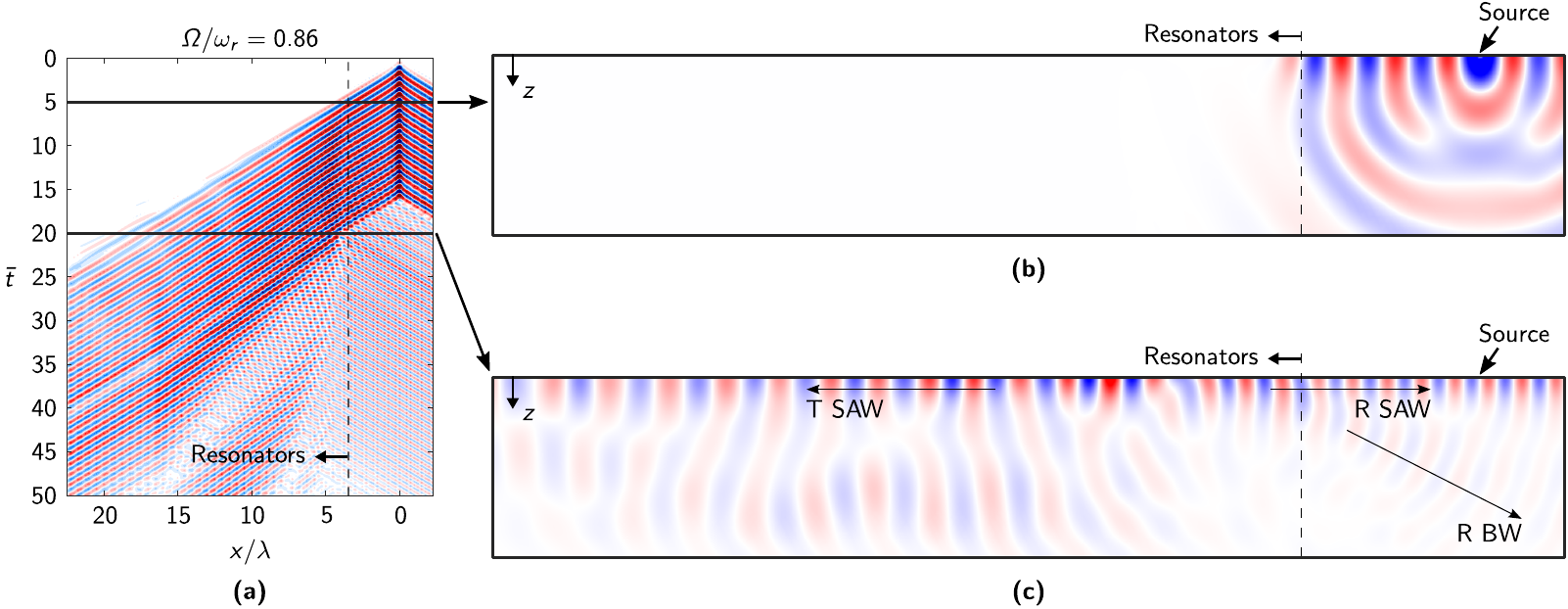}
\caption{(a) Time-space evolution of the surface displacement for a left-propagating wave at $0.86\,\omega_r$. The dashed line indicates the beginning of the region that features resonators. The thick horizontal lines indicate the time instants of interest. (b) The wavefield at $\bar{t}=5$, showing how waves propagate along and below the surface. (c) The wavefield at $\bar{t}=20$. The arrows and letters indicate wave features of interest.}
\label{f:WF}
\end{figure*}

It is known that surface waves can convert into bulk waves upon interaction with a metasurface~\cite{Colquitt2017}. To evaluate how this phenomenon can influence the directional response of a modulated metasurface,  we analyze the full wavefield in the substrate at different time instants. We consider the case of a left-propagating narrow-band signal with carrier frequency $\Omega=0.86\,\omega_r$. This case corresponds to the results in Fig.~\ref{f:TR}(i,j). The time-space evolution of the displacement field along the surface is illustrated in Fig.~\ref{f:WF}(a). 
The wavefields corresponding to time instants $\bar{t}=5$ and $\bar{t}=20$ are shown in Fig.~\ref{f:WF}(b,c), respectively. In particular, the wavefield at $\bar{t}=20$ presents several interesting features. First, it is clearly visible that the transmitted and reflected surface waves have different wavelength contents, as a result of the frequency conversion shown in Fig.~\ref{f:TR}(i). This is an example of conversion by reflection due to spatio-temporal modulations~\cite{Nassar2017eml}. The conversion does not take place exactly at the edge of the resonators region, but rather at a location within the resonator array.  
If we focus our attention on the reflected waves, we can also see that not all waves are reflected along the surface. As indicated by the arrow pointing towards the bottom-right of Fig.~\ref{f:WF}(c), a part of the scattered field is converted into waves that propagate towards the bulk. It would be interesting to quantify the surface-to-bulk wave conversion mechanism and determine the penetration length of the fundamental wave into the metasurface. These aspects, which have practical implications for the design of surface wave converters and filters, deserve a separate treatment.

\section{Conclusions}
\label{s:concl}




We have provided a detailed analytical and numerical account of the non-reciprocal propagation of surface waves of the Rayleigh type in a dynamically modulated metasurface. We have first bridged the gap between the single-resonator dynamics and wave-resonator interactions, by providing a detailed description of the dynamics of a time-modulated resonator. We have then developed an analytical framework to describe the dispersion properties of spatio-temporally varying metasurfaces, and illustrated their asymmetric features.
By means of numerical simulations, we have demonstrated the occurrence of non-reciprocal surface wave attenuation, frequency conversion by reflection and by transmission. We have also shown that surface waves interacting with the modulated metasurface can leak as bulk waves into the substrate. Our findings and the tools we have provided can serve as guidelines for future experiments on the topic, and can play an important role in developing practical designs of SAW devices with unprecedented wave manipulation capacity.

\section*{Acknowledgments}
AP acknowledges the support of DICAM at the University of Bologna. PC acknowledges the support of the Research Foundation at Stony Brook University. CD acknowledges support from the National Science Foundation under EFRI Grant No.\ 1741565. The authors wish to thank Lorenz Affentranger and Yifan Wang for useful discussions.

\appendix
\setcounter{figure}{0} 

\section{Details on the evaluation of the first-order correction to the surface wave dispersion}
\label{a:analy}
 
\begin{figure*}[!htb]
\centering
\includegraphics[scale=1]{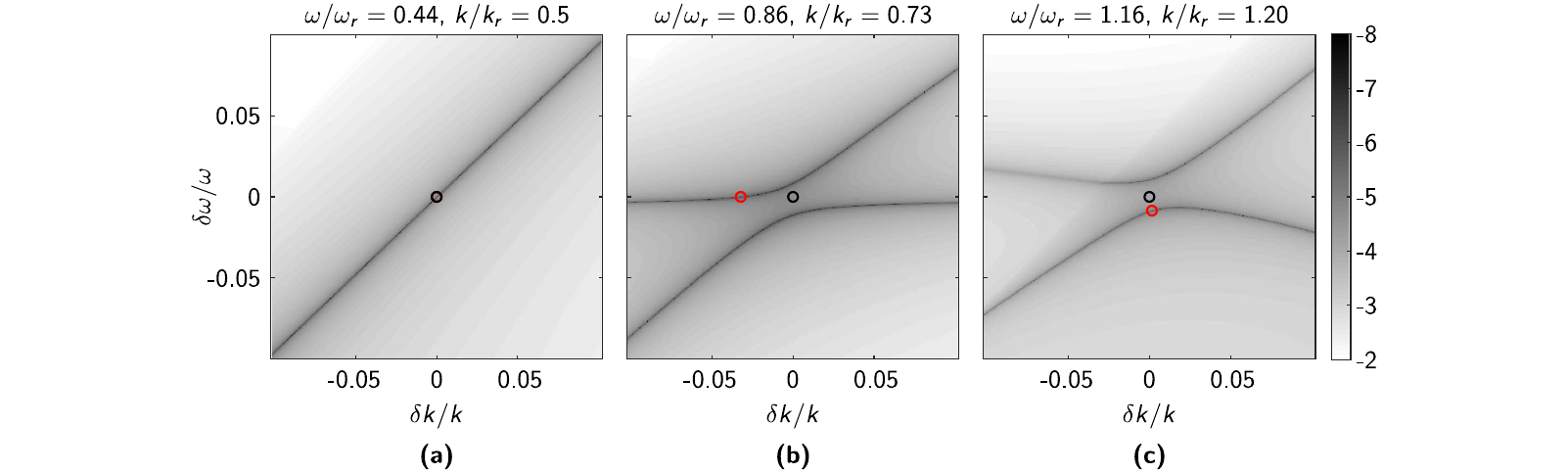}
\caption{Evaluation of the determinant of the matrix in Eq.~\eqref{e:metasurf correction} at (a) a point that does not correspond to a crossing between the fundamental and shifted dispersion curves in Fig.~\ref{f:disp}(a); (b) the veering point $(\omega,k)=(0.86\,\omega_r,0.73\,k_r)$ in Fig.~\ref{f:disp}(b); (c) the locking point $(\omega,k)=(1.16\,\omega_r,1.20\,k_r)$ in Fig.~\ref{f:disp}(c). All cases correspond to a stiffness modulation amplitude $dK=0.05\,K_0$. The colormap bar shows the logarithm (in base 10) of the determinant. The black circular marker denotes $(\delta \omega , \delta k)=(0,0)$. The red circular markers indicate the chosen corrections $\delta k,\delta \omega$.}
\label{f:delta}
\end{figure*}

In this Appendix, we describe the procedure we use to calculate the corrections $\delta k$ and $\delta \omega$ to the fundamental dispersion curve in Section~\ref{s:modmetasurf}. 
First, we compute the determinant of the matrix in Eq.~\eqref{e:metasurf correction} in the vicinity of the  solution  $(\omega,k)$ belonging to the fundamental dispersion curves, within the frequency range $-0.1\omega \leq \delta \omega\leq 0.1\omega$ and wavenumber range  $-0.1k \leq \delta k\leq 0.1k$. 
An example of the resulting amplitude maps is shown in Fig.~\ref{f:delta}(a) for a point located far from any crossing, for the veering point in Fig.~\ref{f:disp}(b), and for the locking in Fig.~\ref{f:disp}(c). The colormaps indicate that multiple locations minimize the determinant of Eq.~\eqref{e:metasurf correction}.
Our procedure for selecting ($\delta \omega$, $\delta k$) is to look for a correction with $\delta \omega=0$ and minimal $\delta k$. This procedure works at veering points, i.e., where the branches interact without opening a frequency gap. 
For directional bandgaps generated at locking crossing points, a solution with $\delta \omega=0$ does not exist. In this case, we choose a correction point with minimal distance from the origin, i.e., $\min\left(\sqrt{{\delta \omega}^2+{\delta k}^2}\right)$.

\section{Conversion by transmission}
\label{a:transm}

\begin{figure*}[!htb]
\centering
\includegraphics[scale=1]{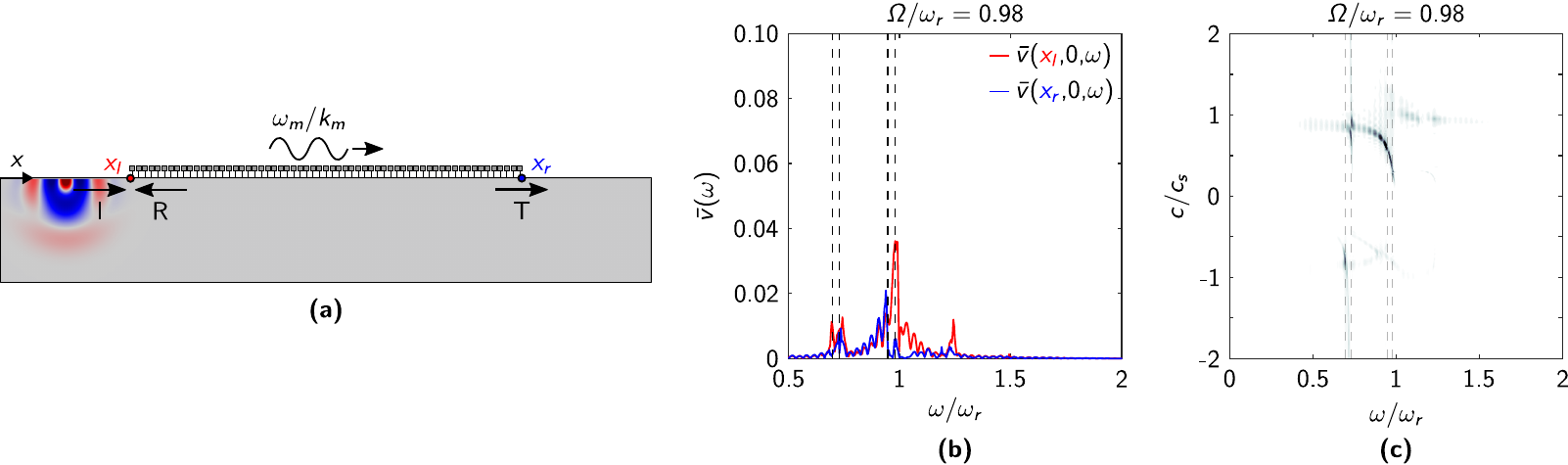}
\caption{Transient FE simulations of a narrow-band signal centred at the veering frequency $0.98\,\omega_r$, and traveling through a modulated metasurface. (a) Schematic of the numerical setup for a right-propagating surface wave. (b) Spectral content of the vertical surface wave field recorded at the left and right edges of the array of resonators. (c) Radon transform of the time-space surface wave recorded along the array of resonators.}
\label{f:TR2}
\end{figure*}

In this Appendix, we briefly describe what happens when the incident wave is centered at a frequency that corresponds to one of the veering points, where co-directional branches of the dispersion repel each other without opening a bandgap.
To do so, we consider a modulating wave with  $dK=0.1$, $\omega_m=0.25\,\omega_r$ and $k_m=2.5\,k_r$. The dispersion relation of this metasurface is shown in Fig.~\ref{f:disp_num}(d). We concentrate on the veering point at $(\omega,k)=(0.98\,\omega_r, 3.36\,k_r)$. This crossing point is coupled to another veering point at $(\omega,k)=(0.73\,\omega_r,0.86\,k_r)$. The modulated surface is excited by a right-propagating wave packet with a frequency centered at $\Omega=0.98\,\omega_r$. As illustrated in Fig.~\ref{f:TR2}(a), the modulating wave is also right-propagating.
In Fig.~\ref{f:TR2}(b), we report the spectral content of the signals recorded at $x_l$ and $x_r$, i.e., at the left edge (red line) and right edge (blue line) of the array of resonators, respectively. Due to the non-zero bandwidth of the excitation signal, the incident wave also excites the locking point at $(\omega,k)=(0.93\,\omega_r,=1.67\,k_r)$. This crossing point, in turn, is coupled to the locking point at $(\omega,k)=(0.69\,\omega_r,-0.81\,k_r)$. Neither of the locking points are associated with a significantly-wide bandgap.

Since all crossing points of interest are below the resonant bandgap of the system, i.e., below $\omega/\omega_r=1$, we focus on this range of frequencies. The transmitted signal, recorded at $x_r$, indicates peaks of comparable amplitude at $\omega/\omega_r=0.73$ and $\omega/\omega_r=0.98$. This is evidence of a conversion-by-transmission phenomenon typical of veering points: part of the transmitted signal is converted to the frequency that is coupled to the $\omega=0.98\,\omega_r$ veering point. In fact, The Radon transform in Fig.~\ref{f:TR2} indicates that the wave associated to the peak at $\omega=0.73\,\omega_r$ has a positive velocity.
The signal at $x_r$ also shows a peak near $\omega=0.93\,\omega_r$, caused by the excitation of the locking point.
The incident signal, recorded at $x_l$, presents peaks at $\omega=0.98\,\omega_r$, $\omega=0.69\,\omega_r$ and $\omega=0.73\,\omega_r$. The Radon transform in Fig.~\ref{f:TR2} indicates that the peak at $\omega=0.69\,\omega_r$, being associated to the locking point at $\omega=0.93\,\omega_r$, has a negative velocity.

\bibliographystyle{elsarticle-num}

\end{document}